# Domain-Adversarial Transfer Learning for Fault Root Cause Identification in Cloud Computing Systems

Bruce Fang
Cornell University
Ithaca, USA

Danyi Gao*
Columbia University
New York, USA

*Abstract-This paper addresses the challenge of fault root cause identification in cloud computing environments. The difficulty arises from complex system structures, dense service coupling, and limited fault information. To solve this problem, an intelligent identification algorithm based on transfer learning is proposed. The method introduces a shared feature extraction module and a domain adversarial mechanism to enable effective knowledge transfer from the source domain to the target domain. This improves the model's discriminative ability and generalization performance in the target domain. The model incorporates a pseudo-label selection strategy. When labeled samples are lacking in the target domain, high-confidence predictions are used in training. This enhances the model's ability to recognize minority classes. To evaluate the stability and adaptability of the method in real-world scenarios, experiments are designed under three conditions: label scarcity, class imbalance, and heterogeneous node environments. Experimental results show that the proposed method outperforms existing mainstream approaches in several key metrics, including accuracy, F1-Score, and AUC. The model demonstrates stronger discriminative power and robustness. Notably, under extreme class imbalance and significant structural differences in the target domain, the model still maintains high performance. This validates the effectiveness and practical value of the proposed mechanisms in complex cloud computing systems.*

I. Introduction

In the context of accelerating digital transformation, cloud computing has become a core pillar of information technology and is widely applied in government, enterprises, and various service platforms. Cloud systems manage and allocate large-scale computing and storage resources in a centralized service-oriented manner, significantly improving resource utilization and service efficiency[1]. However, as the structure of cloud computing systems becomes increasingly complex, the risk of potential failures during operation continues to grow. Due to intricate service dependencies within cloud environments, any abnormality in one component may trigger system-wide cascading effects, severely affecting service availability and user experience. Therefore, enhancing the capability of failure detection and diagnosis, especially the accurate and rapid identification of root causes, is essential for ensuring system stability and reliability[2].

Traditional fault diagnosis methods in cloud computing rely on expert knowledge and rule-based systems. These methods struggle to handle the diverse, dynamic, and log-intensive nature of cloud environments. They often suffer from limited detection accuracy, poor generalization, and insufficient capability to address novel faults. As system runtime increases, multi-source heterogeneous monitoring data accumulate at an exponential rate. This creates an opportunity to improve fault diagnosis through data-driven approaches [3]. Against this background, intelligent fault diagnosis using machine learning has become a research focus. Deep learning, in particular, has improved efficiency and accuracy through automatic feature extraction and pattern recognition [4]. However, deep models require large amounts of labeled data for training, and in cloud systems, real-world fault data—especially representative root cause samples—are scarce. This limits the scalability of deep learning models in practical scenarios.

To address the conflict between data scarcity and limited generalization, transfer learning provides a promising solution. As a machine learning technique that transfers knowledge from related domains to assist the target task [5], transfer learning enables more effective root cause identification in cloud systems. Its core idea is to apply knowledge learned from the source domain to the target domain, reducing the reliance on large-scale labeled data and improving learning efficiency under few-shot conditions[6]. In cloud computing, although systems and platforms vary, their underlying failure mechanisms and operation patterns often exhibit similarities. Transfer learning leverages these commonalities to adapt and transfer diagnostic capabilities across systems. This enhances both the adaptability and generalization of the models.

Transfer learning-based root cause identification not only alleviates the problem of insufficient labeled data but also improves the real-time performance and accuracy of diagnostics. It holds significant value for building intelligent cloud operations and maintenance systems. On one hand, it reduces manual intervention and maintenance costs, supporting a higher degree of automation. On the other hand, knowledge sharing across systems enables continuous optimization of cloud services[7]. As cloud computing evolves toward heterogeneous, edge, and multi-cloud architectures, the complexity of system environments continues to rise. Traditional methods face growing challenges, while the flexibility and adaptability of transfer learning offer strong potential for future cloud fault diagnosis applications[8].

Therefore, in addressing root cause identification in cloud systems, it is important to explore the mechanisms and technical methods of transfer learning in this domain. Such research holds both theoretical and practical value. It promotes the advancement of intelligent diagnostic technologies and supports the development of reliable and highly available cloud service platforms. Building more intelligent, efficient, and transferable models will lay a solid foundation for future automated cloud operations, enhancing the quality of IT infrastructure services and ensuring business continuity.

## II. RELATED WORK

Recent advances in root cause identification for cloud computing systems have been significantly influenced by developments in transfer learning, deep learning, and intelligent system modeling. As systems grow in scale and complexity, traditional rule-based fault analysis techniques increasingly fall short in providing real-time and accurate diagnostics. Hence, leveraging data-driven methodologies, particularly transfer learning, has emerged as a promising direction.

Transfer learning is central to the challenge of adapting fault diagnosis models across different cloud environments. Panigrahi et al. provide a detailed survey of transfer learning techniques, laying a foundational understanding for applying such strategies in dynamic cloud scenarios [9]. Sufian et al. further extend this by exploring deep transfer learning in edge computing, highlighting its utility in mitigating domain shifts and handling data scarcity through cross-domain knowledge reuse [10]. Building on these principles, Rossi et al. integrate transfer learning into workload forecasting for cloud systems, focusing on uncertainty-aware predictions that enhance model robustness in dynamic workloads [11].

Complementary to transfer learning, meta-learning frameworks such as the one proposed by Tang offer robust adaptability for elastic scaling across services, which parallels the model generalization goals in fault diagnosis [12]. These methods help optimize cloud operations in response to fluctuating demands, aligning with the current study's emphasis on transferability and low-resource adaptability. Deep learning also plays a crucial role in managing the complexity of cloud system data. Xin and Pan introduce a self-attention-based model for predicting system performance trends using multi-source metrics, enhancing temporal feature extraction for fault prediction tasks [13]. Similarly, Wang et al. apply time-series deep neural structures to enable proactive fault prediction in distributed systems, demonstrating the efficacy of deep learning in real-time cloud diagnostics [14].

Reinforcement learning techniques have further enriched fault tolerance strategies. Duan explores a TD3-based approach for continuous control in load balancing, providing insights into adaptive learning for performance optimization [15]. Sun et al. adopt a Deep Q-Network framework for intelligent cache management, showcasing deep reinforcement learning's role in backend system efficiency [16].

In broader AI contexts, structural and multi-task learning techniques offer transferable methodologies relevant to cloud diagnostics. Xing proposes bootstrapped structural prompting for analogical reasoning in language models, presenting a novel approach to model generalization and few-shot learning [17]. In tandem, Zhang et al. develop a unified multi-task learning strategy with gradient coordination, reinforcing the value of shared feature spaces for handling diverse learning tasks [18].

Collectively, these studies reflect a growing trend toward adaptive, data-efficient, and generalizable models in cloud operations. By integrating transfer learning with robust deep learning strategies, the field advances toward more resilient and intelligent cloud computing diagnostics.

## III. PROPOSED METHODOLOGY

This study proposes a cloud computing fault root cause identification method based on transfer learning, aiming to solve the problems of scarce labeled data and insufficient model generalization in the target domain. This method obtains representative feature expressions from the source domain and realizes knowledge transfer and identification of target domain fault data through a migration mechanism.

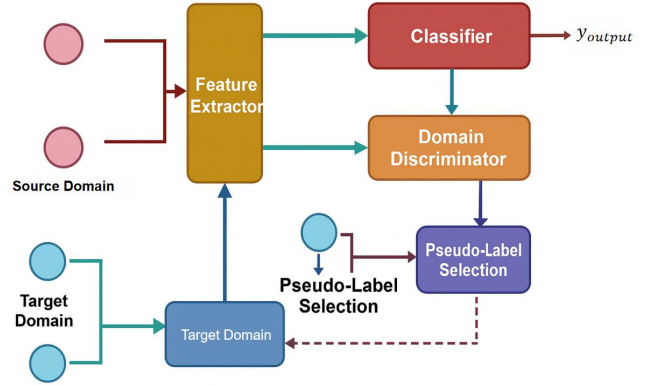

Figure 1. Architecture of the Transfer Learning–Based Fault-Root-Cause Identification Model

Assume that the source domain dataset is $D_s = \{(x_i^s, y_i^s)\}_{i=1}^{n_s}$ and the target domain unlabeled data is $D_t = \{x_j^t\}_{j=1}^{n_t}$, where x represents the monitoring data feature vector and y represents the fault category label. By jointly training the source domain and target domain features, the model has good discrimination ability in the source domain and adaptability to the target domain.

In the model design, a shared feature extractor $F(\cdot; \theta_f)$ is first constructed to extract high-dimensional semantic features of the input data. Subsequently, the classifier $C(\cdot; \theta_c)$ is used on the source domain to classify the extracted features and minimize the classification loss function of the source domain:

$$L_s = \frac{1}{n_s} \sum_{i=1}^{n_s} l(C(F(x_i^s)), y_i^s) \quad (1)$$

Where $l(\cdot)$ represents the cross entropy loss function. To enable the model to learn effective features in the target domain, this paper introduces a domain alignment mechanism and adopts the Maximum Mean Discrepancy (MMD) method to minimize the distance between the feature distributions of the source domain and the target domain, which is defined as:

$$L_{mmd}=\| \frac{1}{n_s}\sum_{i=1}^{n_s} F(x_i^s) - \frac{1}{n_t}\sum_{i=1}^{n_t} F(x_j^t) \|^2 \quad (2)$$

In order to further improve the domain adaptability of the model, an adversarial training mechanism is introduced. By introducing a domain discriminator $D(\cdot;\theta_d)$, it is possible to discriminate whether the input features come from the source domain or the target domain. The feature extractor is trained through an adversarial loss function so that the extracted features are domain-indistinguishable, thereby achieving inter-domain feature alignment. The adversarial loss is defined as follows:

$$L_{adv}= \frac{1}{n_s}\sum_{i=1}^{n_s} \log D(F(x_i^s)) - \frac{1}{n_t}\sum_{i=1}^{n_t} \log(1-D(F(x_j^t))) \quad (3)$$

Among them, the feature extractor parameter $\theta_f$ is updated by minimizing $L_{adv}$, and the domain discriminator parameter $\theta_d$ is updated by maximizing the loss, thereby achieving the game-based optimization goal. During the entire training process, the model is optimized end-to-end using a joint loss function:

$$L_{total} = L_s + \lambda_1 L_{mmd} + \lambda_2 L_{adv} \quad (4)$$

Among them, A and B are the weight coefficients for balancing various losses.

In addition, to further improve the classification performance of the target domain, this paper introduces a pseudo-label mechanism. By predicting the target domain data and selecting samples with high confidence as pseudo-labels, they are added to the source domain data to participate in the model update. Let the target domain prediction label be $\hat{y}_j^t = \arg\max C(F(x_j^t))$. When the prediction confidence is $p(\hat{y}_j^t) \geq \delta$, the sample is selected as a pseudo-labeled sample to participate in the semi-supervised training process. This pseudo-label enhancement mechanism further alleviates the problem of target domain data scarcity, while improving the model's generalization ability and robustness to actual cloud platform failures. The overall structure of this method is both stable and scalable and is suitable for a variety of cloud computing operation and maintenance scenarios.

IV. EMPIRICAL EVALUATION

A. Dataset Description

This study uses the publicly available Alibaba Cluster Trace 2018 dataset as the primary data source. The dataset consists of real operational data from a large-scale cloud computing platform. It includes rich records of resource usage, task scheduling information, and node-level runtime logs. The data covers multidimensional monitoring information from thousands of servers, including key performance indicators such as CPU usage, memory consumption, disk read and write activity, and network load. These features provide strong support for fault detection and root cause analysis.

In this study, the source and target domains are defined based on operational characteristics across different periods and groups of nodes. The source domain data is selected from a set of stable nodes operating under high load, which contains a large number of labeled fault events and clearly defined fault types. This makes it suitable for model training and knowledge extraction. The target domain data is taken from another group of nodes that differ in structure or runtime environment. These logs contain partially labeled or unlabeled data, closely reflecting real-world conditions where fault labels are often incomplete or unavailable. This setup effectively simulates the cross-environment knowledge transfer challenges in transfer learning.

The selection of this dataset ensures both authenticity and diversity. It also reflects the multi-source heterogeneity and temporal dynamics typical of cloud computing environments. Using this dataset for modeling helps evaluate the adaptability and discriminative ability of the algorithm in real-world scenarios. At the same time, it provides a reliable data foundation for building fault root cause identification models that are both generalizable and transferable.

B. Experimental Results

This paper first gives the comparative experimental results, as shown in Table 1.

As shown in the experimental results table, the proposed method achieves the highest accuracy (89.6%) in cloud computing fault root cause identification, demonstrating superior discriminative capability in complex environments. It also leads in F1-Score (85.4%), indicating effective balance between precision and recall under conditions of class imbalance and limited fault samples. Furthermore, it attains the top AUC score (91.3%), reflecting strong robustness and generalization in distinguishing fault types across heterogeneous domains. These results highlight the advantages of integrating domain alignment, pseudo-labeling, and adversarial learning, which together enhance adaptability in label-scarce or noisy settings. Compared with mainstream methods such as DANN, CDAN, FixBi, and ToAlign, the proposed model consistently outperforms across all metrics, validating its effectiveness and scalability for intelligent cloud operations. A robustness analysis under limited-label conditions is further illustrated in Figure 2.

Table 1. Comparative experimental results

| Method | Accuracy | F1-Score | AUC |
|---|---|---|---|
| DANN[19] | 84.2 | 81.0 | 87.2 |
| CDAN[20] | 85.7 | 82.3 | 88.5 |
| FixBi[21] | 86.3 | 83.1 | 89.2 |
| Toalign[22] | 87.1 | 83.7 | 89.9 |
| Ours | 89.6 | 85.4 | 91.3 |

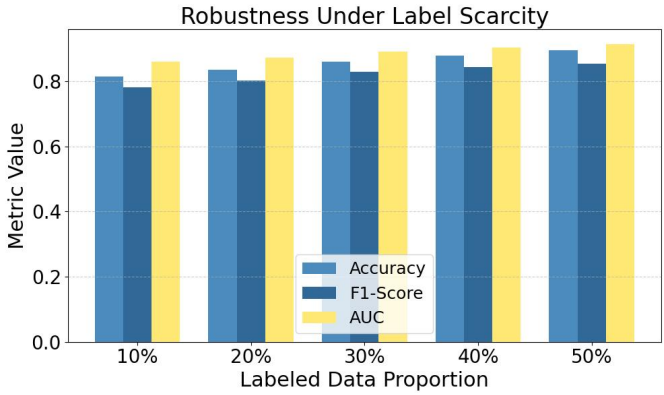

Figure 2. Robustness analysis of transfer models in label-scarce scenarios

The figure illustrates that the proposed transfer learning model demonstrates a consistent improvement in accuracy, F1-Score, and AUC as the proportion of labeled data increases, confirming its robustness and adaptability under extreme label scarcity. Remarkably, with only 10% labeled data, the model achieves an accuracy of 0.81 and an AUC of 0.86, indicating effective knowledge transfer and fault feature recognition through pseudo-labeling and domain adaptation. The steady rise in F1-Score further reflects balanced performance across imbalanced classes, maintaining precision and recall even with limited target samples. These results underscore the model's suitability for real-world, low-label scenarios in cloud operations. Its stability is attributed to the integration of joint feature extraction, domain alignment, and pseudo-label filtering. A comparison of model performance under imbalanced fault category conditions is presented in Figure 3.

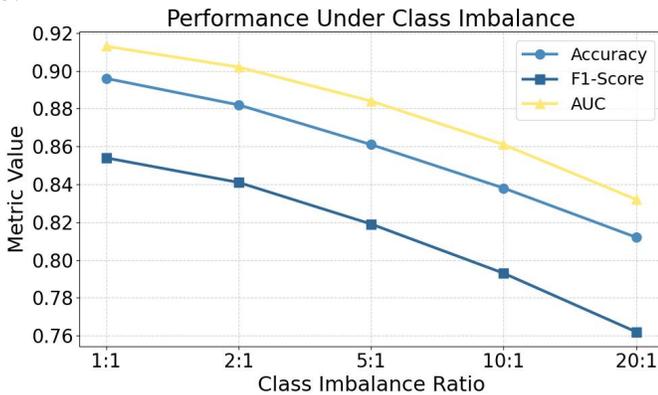

Figure 3. Comparison of performance of various models under the condition of imbalanced fault categories

The trend illustrated in the figure shows that the model's performance declines across all metrics as fault class imbalance increases, indicating a reduced ability to detect minority class faults under skewed distributions. This degradation is particularly relevant in cloud computing environments, where rare but critical faults can significantly impact system stability. The continued drop in F1-Score suggests a worsening balance between precision and recall, reflecting challenges in maintaining classification boundaries and highlighting the limitations of transfer feature learning under label-scarce, imbalanced conditions. Although AUC remains relatively high, its downward trend further indicates that class imbalance hampers the model's ability to distinguish positive and negative samples, especially in multi-source heterogeneous settings. These findings underscore the importance of addressing distribution sensitivity when designing robust transfer learning models. In real-world scenarios with inherently imbalanced data, transfer mechanisms alone may be insufficient; strategies such as sample reweighting and improved pseudo-label selection are needed to enhance reliability. An evaluation of the model's adaptability using a domain adversarial mechanism on heterogeneous nodes is presented in Figure 4.

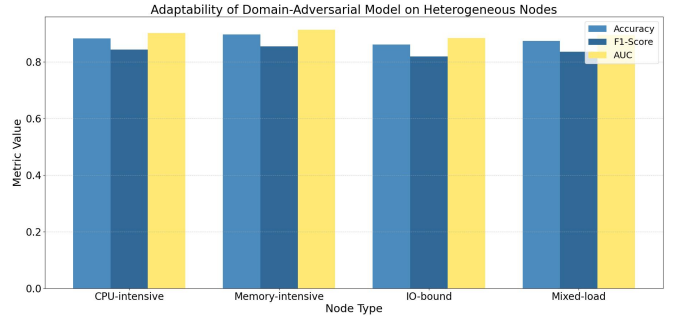

Figure 4. Evaluation of the adaptability of models with domain adversarial mechanisms on heterogeneous nodes

The figure demonstrates the adaptability of the proposed transfer learning model with domain adversarial mechanisms across various heterogeneous node types. Overall, the model maintains high performance on all metrics, whether on compute-intensive nodes or those characterized by memory, I/O, or mixed workloads. This demonstrates strong cross-platform generalization and stability. The results confirm the effectiveness of the domain adversarial strategy in reducing distributional differences among different node types. On memory-intensive and CPU-intensive nodes, the model shows particularly high accuracy and discriminative ability. This suggests that the operational patterns of these nodes are more stable, making it easier for the domain adversarial mechanism to capture transferable features. This has practical relevance for cloud platforms with clearly partitioned resource pools. It supports proactive resource scheduling and fault prediction.

For I/O-bound and mixed-load nodes, although performance shows slight fluctuations, it remains at a high level overall. This indicates that the model has a certain level of robustness in handling dynamic and mixed-load scenarios. It can effectively adapt to the high variability commonly seen in real-world cloud environments. The domain adversarial mechanism plays a key role in this process by enabling the feature extractor to generate unified representations that are domain-invariant. In summary, this experiment further validates the practical value of the proposed method in diverse and complex cloud architectures. By achieving robust transfer classification across heterogeneous nodes, the model enhances intelligent operations and maintenance in the presence of diverse hardware infrastructures. This supports the development of a more generalizable solution for cloud fault root cause identification.

## V. Conclusion

This study addresses key challenges in fault root cause identification within cloud computing environments, including high-dimensional complexity, label scarcity, and system heterogeneity. It proposes an intelligent identification method based on transfer learning. The method uses a shared feature extractor and domain adversarial structure to enable effective knowledge transfer from the source domain to the target domain. This significantly mitigates the problem of insufficient labeled data in the target domain. Additionally, a pseudo-label selection mechanism is integrated to enhance the model's adaptability, improving fault recognition accuracy and robustness in real-world cloud platforms. The approach offers a practical path toward building automated operations and maintenance systems.

Overall, the proposed model demonstrates strong performance in accuracy, F1-Score, and AUC. It maintains a high discriminative ability even under challenging conditions such as limited labels, class imbalance, and heterogeneous nodes. This confirms the effectiveness and generalizability of the transfer strategy and model design when applied to real cloud computing data. Furthermore, the experimental results highlight the benefits of the coordinated function of different components. They validate the positive contribution of domain adversarial learning and pseudo-label enhancement in improving recognition in the target domain.

The outcomes of this study enrich the theoretical foundation of transfer learning in cloud fault diagnosis and provide practical insights for deploying intelligent operations systems. In typical cloud scenarios such as multi-tenancy, high dynamism, and low-label availability, the proposed method can serve as a core module within existing monitoring and maintenance platforms. It enhances system-level awareness and proactive response. The method also shows strong engineering adaptability and practical value in areas such as enterprise-scale distributed systems, large-scale cluster management, and edge node diagnostics.

## VI. Future Research Directions

Future work may explore several directions. One is to incorporate graph neural networks or attention mechanisms to improve the modeling of complex dependencies and temporal features. Another is to integrate multimodal data, including logs, metrics, and alerts, to construct a more generalizable decision framework. Additionally, practical deployment issues such as online transfer, incremental learning, and model compression deserve further investigation. These efforts will help promote the full-scale adoption and efficient implementation of transfer learning in intelligent cloud operations.